\documentclass{article}          

\usepackage{graphicx}
\usepackage{amssymb}
\usepackage{amsmath}
\usepackage{xcolor}
\usepackage{graphicx}
\usepackage{placeins}
\usepackage{csquotes}

\newtheorem{definition}{Definition}

\newcommand{\bit}{\begin{itemize}}
\newcommand{\eit}{\end{itemize}}
\newcommand{\bd}{\begin{description}}
\newcommand{\ed}{\end{description}}
\newcommand{\ben}{\begin{enumerate}}
\newcommand{\een}{\end{enumerate}}
\newcommand{\bqn}{\begin{equation*}\begin{aligned}}
\newcommand{\eqn}{\end{aligned}\end{equation*}}
\newcommand{\bqnn}{\begin{equation}\begin{aligned}}
\newcommand{\eqnn}{\end{aligned}\end{equation}}
\newcommand{\bt}{\begin{thm}}
\newcommand{\et}{\end{thm}}
\newcommand{\bl}{\begin{lem}}
\newcommand{\el}{\end{lem}}
\newcommand{\bp}{\begin{prop}}
\newcommand{\ep}{\end{prop}}
\newcommand{\bc}{\begin{cor}}
\newcommand{\ec}{\end{cor}}
\newcommand{\bdefn}{\begin{defn}}
\newcommand{\edefn}{\end{defn}}
\newcommand{\brem}{\begin{rem}}
\newcommand{\erem}{\end{rem}}
\newcommand{\bproof}{\begin{proof}}
\newcommand{\eproof}{\end{proof}}
\newcommand{\bex}{\begin{ex}}
\newcommand{\eex}{\end{ex}}
\newcommand{\R}{\mathbb{R}}

\newcommand{\bcs}{\begin{cases}}
\newcommand{\ecs}{\end{cases}}

\newcommand{\expp}{\text{e}}

\newcommand{\se}{\text{SE}(2)}

\newcommand{\INT}{\int\limits}
\newcommand{\SUM}{\sum\limits}

\newcommand{\omg}{\omega}
\newcommand{\abs}[1]{\lvert {#1} \rvert}

\begin{document}

\title{A sub-Riemannian model of the visual cortex with frequency and phase
}
\date{}
%
\author{E. Baspinar\thanks{INRIA Sophia Antipolis, MathNeuro Team (corresponding author), emre.baspinar@inria.fr } \and A. Sarti\thanks{EHESS, CAMS, alessandro.sarti@ehess.fr}\and G. Citti\thanks{University of Bologna, Department of Mathematics, giovanna.citti@unibo.it} \footnotemark[1]}

\maketitle

\graphicspath{{figures/}}

\begin{abstract}
In this paper we present a novel model of the primary visual cortex (V1) based on orientation, frequency and phase selective behavior of the V1 simple cells. We start from the first level mechanisms of visual perception: receptive profiles. The model interprets V1 as a fiber bundle over the 2-dimensional retinal plane by introducing orientation, frequency and phase as intrinsic variables. Each receptive profile on the fiber is mathematically interpreted as a rotated, frequency modulated and phase shifted Gabor function. We start from the Gabor function and show that it induces in a natural way the model geometry and the associated horizontal connectivity modeling the neural connectivity patterns in V1.  We provide an image enhancement algorithm employing the model framework. The algorithm is capable of exploiting not only orientation but also frequency and phase information existing intrinsically in a 2-dimensional input image. We provide the experimental results corresponding to the enhancement algorithm. 

\vspace{0.5cm}



\end{abstract}

\section{Introduction}

The question of how we perceive has been an intriguing topic for different disciplines. 
One of the first school which faced the problem is the Berlin school of experimental psychology, called \emph{Gestalt} \emph{psychology} school, \cite{wertheimer1938laws}, \cite{kohler1970gestalt}, \cite{koffka2013principles}
which formulates precise laws which can explain visual perception. 
The Gestalt psychology 
is a theory for understanding the principles underlying 
the emergence of perceptual units, as the result of a grouping process. 
The main idea is that perception is a global
phenomenon, which considers the scene as a whole, and 
is much more than the pure sum of local perception. 
The first perceptual laws are of qualitative type, based on similarity, closure, good continuation, alignment. 
After that there have been many psychophysical studies which attempted to provide quantitative version of the grouping process. With the developments of neuroscience studies, researchers started to look for cortical implementation of Gestalt laws, with a particular attention to neural architectures of the visual cortex. A particularly important one for our study  is the pioneering work of Field et. al. \cite{field1993contour}, which models Gestalt principles of good continuation and alignment. They experimentally proved that fragments aligned along a curvilinear path can be perceived as a unique perceptual unit much better than fragments with rapidly changing orientations. The results of their experiments 
were summarized in a representation, called \emph{association fields}, which represent the complete set of paths with fixed initial position and orientation which can be perceived as perceptual units. The visual cortex is a part of the mammalian brain which is responsible for the first level processing tasks of perceptual organization of local visual features in a visual stimulus (two dimensional image). It is known from neurophysiological experiments that the visual cortex contains neurons (simple cells) which are locally sensitive to several visual features, namely, orientation \cite{hubel1959receptive}, \cite{hubel1962receptive}, \cite{hubel1963shape}, \cite{hubel1977ferrier}, spatial frequency \cite{maffei1977spatial}, \cite{hubener1997spatial}, \cite{issa2000spatial}, \cite{issa2008models}, \cite{sirovich2004organization}, \cite{tani2012parallel},  \cite{ribot2013organization}, \cite{ribot2016pinwheel},  phase \cite{de1983spatial}, \cite{pollen1988responses}, \cite{levitt1990spatio},  \cite{mechler2002detection}, scale \cite{blakemore1969existence} and ocular dominance \cite{shatz1978ocular}, \cite{levay1978ocular}, \cite{issa2000spatial}. The simple cells are organized in a {hypercolumnar architecture}, which was first discovered by Hubel and Wiesel \cite{hubel1974uniformity}. In this architecture, a hypercolumn is assigned to each point $(x,y)$ of the retinal plane $M\simeq \R^2$ (if we disregard the isomorphic cortical mapping between retinal and cortical planes), and the hypercolumn contains all the simple cells sensitive to a particular value of the same feature type. Simple cells are able to locally detect features of the visual stimulus, and neural connectivity between the simple cells integrates them in a coherent global unity. Those two mechanisms, the feature detection and the neural connectivity, comprise the functional geometry of V1.

Several models were proposed for the functional geometry of V1 associated to the simple cells which were only orientation sensitive. Early models date back to '80s. Koenderink and van Doorn \cite{koenderink1984structure}, \cite{koenderink1987representation} revealed the similarity between Gaussian derivative functions and simple cell receptive profiles. They proposed visual models based on the functions of Gaussian derivatives as the mathematical representations of the receptive profiles. Their findings indeed encouraged many studies relying on the choice of a family of Gaussian derivative functions and Gaussian kernels, among which we would like to mention the works of Young \cite{young1987gaussian} and Lindeberg \cite{lindeberg1998feature}, \cite{lindeberg2013computational}. 

A different modeling approach from the above mentioned ones was to employ Gabor functions as the mathematical representations of the orientation sensitive simple cell receptive profiles. The motivation for this choice was relying on an uncertainty principle as was elaborated by Daugman \cite{daugman1985uncertainty} through a generalization of the hypothesis of Mar{\^{c}}elja \cite{marcelja1980mathematical} (see also \cite{jones1987evaluation} where Jones and Palmer compared statistically the results obtained via Gabor functions and the neurophysiological results collected from V1 of a cat). Furthermore Hoffman (see \cite{hoffman1970higher}, \cite{hoffman1989visual}) proposed to model the hypercolumnar architecture of V1 as a fiber bundle. Following the second school (which uses the Gabor functions) and by further developing the model proposed by Petitot and Tondut \cite{petitot1999vers} (see also \cite{petitot2003neurogeometry} and \cite{petitot2008neurogeometrie} of Petitot), where hypercolumnar architecture was interpreted as a fiber bundle associated to a contact geometry, Citti and Sarti \cite{citti2006cortical} introduced a group based approach. They proposed a new model of the functional geometry of V1, which considered the sub-Riemannian geometry of the roto-translation group ($\se$) as the suitable model geometry. The main reason for employing $\se$ geometry was due to that the corresponding Lie algebra to $\se$ was providing a good model of the actual neural connectivity in V1. This model proposed in \cite{citti2006cortical} has been extended to other visual features in addition to orientation, such as scale by Sarti et. al. \cite{sarti2008symplectic}, and to other cell types such as complex cells sensitive to velocitiy and movement direction by Barbieri et. al. \cite{barbieri2014cortical} and Cocci et. al. \cite{cocci2015cortical}. Apart from those, a semidiscrete model was presented by Prandi et. al. in \cite{prandi2015image}. Furthermore, image processing applications employing Gabor transform in order to extract visual features from medical images were proposed in \cite{duits2013evolution} by Duits and Sharma (see also \cite{sharma2015left}). Other applications in medical image analysis employing scale and orientation information can be found in  \cite{bruurmijn2013myocardial} and \cite{kause2013direct}, where Gabor transform is employed for the detection of local frequencies in tagging MRI (magnetic resonance imaging) images and thus for the computation of local frequency deformations in those images. Interested reader can also refer to \cite{faugeras1993three} for different applications of geometric approach in general, in computer vision and robotics. Additionally to those studies, the models in terms of cortical orientation and orientation-frequency selectivity, which were provided by Bressloff and Cowan \cite{bressloff2003functional}, \cite{bressloff2001geometric}, could be useful references for the reader. We refer to \cite{citti2014neuromathematics} for a review of several cortical models including many of the above mentioned ones. 

The theoretical criterion underpinning the modeling we propose in this paper relies on the so called neurogeometrical approach described by Citti and Sarti \cite{citti2006cortical}, Petitot and Tondut \cite{petitot1999vers}, Sarti et. al. \cite{sarti2008symplectic}. Following this approach, processing capabilities of sensorial cortices, and in particular of the visual cortex are modeled based on the geometrical structure of cortical neural connectivity.
Global and local symmetries of the visual stimuli are captured by the cortical structure which is invariant under those symmetries (see Sanguinetti et. al. \cite{sanguinetti2010model}). We will follow a similar framework and we will start from the first level perceptual tasks performed by the simple cells, from local feature extraction. This starting point will lead us to the model geometry of V1 associated to the simple cells sensitive to orientation, spatial frequency and phase information at each position in a given two dimensional image.  

At the level of Gestalt organisation, the neurogeometrical architecture in $\se$ \cite{citti2006cortical} implements the psychophysical law of good continuation, the architecture in the affine group \cite{sarti2008symplectic} implements good continuation and ladder, the architecture in the Galilean group \cite{barbieri2014cortical}, \cite{cocci2015cortical} implements common fate, the architecture we are considering here in a Gabor based sub-Riemannian geometry implements similarity between textures/patterns and contains all the previous models employing the neurogeometrical approach.

Once the light reflects from a visual stimulus and arrives to the retina, it evokes some spikes which are transmitted along the neural pathways to the simple cells in V1. Each simple cell gives a response called \emph{receptive profile} to those spikes. In other words, receptive profile is the impulse response of a simple cell. The simple cells extract the information of local visual features by using their receptive profiles and it is possible to represent the extracted features mathematically in a higher dimensional space than the given two dimensional image plane. We will call this space \emph{the lifted space} or \emph{the lifted geometry}. We will use an extended Gabor function as the receptive profile of the simple cells. We will see that this choice naturally  induces the corresponding Lie algebra of the sub-Riemannian structure, which is the  corresponding lifted geometry to our model. The Lie algebra and its integral curves model neural connectivity between the simple cells. Moreover, since some pairs of the algebra are not commutative, it is possible to formulate an uncertainty principle and this principle is satisfied by the extended Gabor function. That is, the extended Gabor function minimizes uncertainties arising from simultaneous detection of frequency-phase and simultaneous detection of position-orientation (see also \cite[Section 7.5]{duits2005perceptual}, \cite{barbieri2011coherent}, \cite{barbieri2012uncertainty}, \cite{barbieri2015reproducing} and \cite{sharma2015left}  for similar phenomena in different frameworks).

Concerning the question of which family of functions to use as receptive profiles, let us recall that receptive field models consisting of cascades of linear filters and static non-linearities may be adequate to account for responses to simple stimuli such as gratings and random checkerboards, but their predictions of responses to complicated stimuli (such as natural scenes) are correct only approximately. A variety of mechanisms such as response normalization, gain controls, cross-orientation suppression, intra-cortical modulation can intervene to change radically the shape of the profile. Then any static and linear model for the receptive profiles has to be considered just as a very first approximation of the complex behavior of a real dynamic receptive profile, which is not perfectly described by any of the static wavelet frames.

For example derivatives or difference of Gaussian functions are suitable approximations of the behavior of classical receptive profiles of the simple cells.  In \cite{lindeberg2011generalized, lindeberg2013computational}, Lindeberg starts from certain symmetry properties of the surrounding world and derives axiomatically the functions of Gaussian derivatives obtained from the extension of the family of rotationally symmetric Gaussian kernels to the family of affine Gaussian kernels, and proposes to model the simple cell receptive fields in terms of those Gaussian derivatives (see also Koenderink \cite{koenderink1984structure}, \cite{koenderink1987representation}, Young  \cite{young1987gaussian}, Landy and Movshon \cite{landy1991computational}). Indeed Gaussian functions are good models of the receptive profiles if we restrict ourselves to the visual features except for frequency and phase. They provide good results for orientation and scale detection as shown by the scale-space school  (see, e.g., the works of Lindeberg \cite{lindeberg1994scale}, \cite{lindeberg1998feature}, \cite{lindeberg2013computational}, Florack \cite{florack1997image}, ter Haar Romeny \cite{ter2003front}, \cite{ter2010multi}, Hannink et. al. \cite{hannink2014crossing}). However, we are interested here in two dimensional visual perception based on orientation, frequency and phase sensitive simple cells. Differently from the case with orientation-scale sensitive simple cells, frequency-phase sensitive simple cells cannot be modeled in a straightforward way by Gaussian derivative functions.  A different order Gaussian derivative must be used for the extraction of each frequency component of a given image. This requires the use of different functions of each one of them corresponds to a certain frequency, thus to a certain order derivative. In other words, frequency is not a parameter as in the case of scale but each frequency corresponds to a different function. It is not possible to derive a natural geometry starting from the derivatives of the Gaussian and it is rather required to assign an adequate geometric setting to the set of extracted feature values by the Gaussian derivatives in order to represent those values. At this point, a Gabor function seems to be a good candidate for the detection of different orientation, frequency and phase values in a two dimensional image, since orientation, frequency and phase are parameters of the Gabor function. In other words, instead of using different functions, we can use a single Gabor function corresponding to a set of parameter values in order to detect different feature values. In this way, we obtain a sub-Riemannian model geometry as the natural geometry induced directly by the Gabor function (i.e., by the receptive profile itself).  Moreover, the Gabor function is able to model both asymmetric simple cells and even/odd symmetric simple cells thanks to its phase offset term appearing in its wave content while the functions of the Gaussian derivatives account only for the symmetric simple cells. Considering those points, we propose to use a Gabor function with frequency and phase parameters as the receptive profile model. The Gabor function allows to extend the model provided in \cite{citti2006cortical} to the true distribution of the profiles in V1 (including the asymmetric receptive profiles with the phase shifts) in a straightforward way. Finally, we would like to refer to Duits and Franken \cite{duits2009line}, \cite{duits2010left}, \cite{duits2010left2}, Franken and Duits \cite{franken2009crossing}, Sharma and Duits \cite{sharma2015left}, Zhang et. al. \cite{zhang2016robust}, Bekkers et. al. \cite{bekkers2018roto} for information about applications which employ other wavelets corresponding to unitary transforms for feature extraction. 

Here we consider the model framework provided in \cite{citti2006cortical} as the departure point of our study. We extend this model from orientation selective framework to an orientation, frequency and phase selective framework. Furthermore we provide the neural connectivity among the simple cells not only orientation selective but also frequency selective with different phases. Thanks to the use of all frequency components of the Gabor functions, Gabor transform can be followed by an exact inverse Gabor transform, which was not the case in the model presented in \cite{citti2006cortical} since a single frequency component of the Gabor function was used.    The projection of our generalized model onto $\se$ can be considered as equivalent to the model provided in \cite{citti2006cortical}. The procedure that we use to obtain the extended framework can be employed for the extension to a model associated with orientation-scale selective simple cells as well (see \cite{baspinar2018geometric}).

We will see in Section \ref{sec:Extendeed_model_geometry} the model structure. We will show how the model geometry with the associated horizontal connectivity can be derived starting from the receptive profile model, i.e., from the Gabor function. Then in Section \ref{sec:Connectivity_in_the_extended_phase_space} we will provide the explicit expressions of the horizontal integral curves, which are considered as the models of the association fields in V1. Finally in Section \ref{sec:EnhancementSection}, we will provide an image enhancement algorithm using the model framework together with the results obtained by applying a discrete vesion of the algorithm on some test images.


\section{The model}\label{sec:Extendeed_model_geometry}

The model is based on two mechanisms. The first one is the feature extraction linear mechanism. 
The second mechanism is the propagation along horizontal connectivity, which models the neural connectivity in V1. 
We describe the model by using those two mechanisms in terms of both a group structure and a sub-Riemannian structure.

\subsection{Feature extraction and representation}

\subsubsection{Receptive profiles, symplectic structure and contact form}
\label{sec:receptiveProfilesSymplectic}

Being inspired by the receptive profile models proposed in \cite{citti2006cortical} for the orientation selective behavior and in \cite{deangelis1993spatiotemporal}, \cite{cocci2012spatiotemporal},  \cite{barbieri2014cortical} for the spatio-temporal behavior of the simple cells , we propose to represent the receptive profile of a simple cell in our setting with the Gabor functions of the type
\begin{equation}\label{eq:gaborExtendedFunctionExtended}
\Psi_{\alpha}(x,y,s):=\expp^{-i\big(r \cdot (x-q_1,\, y-q_2)-v(s-\phi)\big)}\expp^{-\abs{x-q_1}^2-\abs{y-q_2}^2},
\end{equation}
with the spatial frequency\footnote{Spatial frequency refers to $\omg=\frac{2\pi}{\lambda}$ with a wavelength $\lambda>0$ in our terminology.} $\omg>0$ and $r=(r_1, r_2)=(-\omg\sin\theta,\,\omg\cos\theta)$, where we represent the coordinates associated to a 6-dimensional space $\mathcal{N}$ with $\alpha=(q_1, q_2,\phi, r_1, r_2, v)\in \R^6$. 


In the case of V1 complex cells with spatio-temporal dynamics, the variable $v$ represents the velocity of a two-dimensional plane wave propagation (see Barbieri et. al \cite{barbieri2014cortical} for details). However, we are not interested in the complex cells and any temporal behavior, and we can choose $v=1$. In our framework we interpret $s-\phi$ as the phase centered at $\phi$, 

In this way we obtain a  5-dimensional space $\mathcal{M}$
\begin{equation}
\mathcal{M}=\mathbb{R}^2\times S^1\times \R^+\times S^1 \ni \alpha=\{q_1, q_2,\theta,\omega,\phi\}=(q,z),
\end{equation}
where $z$ denotes the feature variables $(\theta,\omg,\phi)\in S^1\times \R^+\times S^1$.
Then we may write the associated Gabor function which is centered at $q\in M$ and sensitive to feature values $z$ by using \eqref{eq:gaborExtendedFunctionExtended} as follows:
\begin{equation}\label{eq:gaborExtendedFunction2}
\Psi_{(q,z)}(x,y,s):=\expp^{-i\big(\omg(-\sin\theta,\,\cos\theta ) \cdot (x-q_1,\, y-q_2)-(s-\phi)\big)}\expp^{-\abs{x-q_1}^2-\abs{y-q_2}^2}.
\end{equation}


The standard Liouville form $r_1dx+r_2dy - v ds$ reduces to 
\begin{equation}\label{eq:one_form_theta}
\Theta_{(\theta,\omg)}=r_1 dx+r_2 dy-ds=-\omg\sin\theta dx+\omg\cos\theta dy-ds.
\end{equation}
Indeed $\Theta$ is a contact form since
\begin{equation}
\Theta\wedge d\Theta \wedge d\Theta = \omg\; dx \wedge dy \wedge d\theta \wedge d\omg \wedge ds,
\end{equation}
is a volume form. In other words it is maximally non-degenerate and it does not vanish at any point on the manifold $\mathcal{M}$.

\subsubsection{Set of receptive profiles}\label{sec:ExtendedsetOfReceptiveProfiles}

An important property of Gabor functions is that they are invariant under certain symmetries. Therefore any Gabor function can be obtained from a reference Gabor function (mother Gabor function), up to a certain transformation law.
 
Let us denote the origin for the layer of a frequency $\omega$ by $0_\omega=(0,0,\omega,0)\in \mathcal{M}$. Then a suitable choice of the mother Gabor function with the frequency $\omg$ is
\begin{equation}\label{eq:extendedMotherGaborFcn}
\Psi_{0_\omega}(x,y,s)=\expp^{-i(\omg y-s )}\expp^{-x^2-y^2}.
\end{equation}
We set
\begin{equation}\label{eq:coordinateTransformLocalGlobal}
A_{(q,\theta,\phi)}(\tilde{x},\tilde{y},\tilde{s})=\begin{pmatrix}
q_1 \\ q_2 \\ \phi
\end{pmatrix}+
\begin{pmatrix}
\cos\theta & -\sin \theta & 0\\
\sin\theta & \cos\theta & 0\\
0 & 0 & 1
\end{pmatrix}\begin{pmatrix}
\tilde{x} \\ \tilde{y} \\ \tilde{s}
\end{pmatrix}=(x,y,s),
\end{equation}
which describes at each frequency the relation between a generic receptive profile centered at $z=(q,\theta,\omega,\phi)$ and the mother Gabor function through
\begin{equation}\label{eq:generalGaborFromMother}
\Psi_{(q,z)}(x,y,s)=\Psi_{0_\omega}\big(A^{-1}_{(q,\theta,\phi)}(x,y,s)\big).
\end{equation}

The set of all receptive profiles obtained from the mother Gabor function with all possible combinations of feature values at each possible frequency is called the \emph{set of receptive profiles}.

\subsubsection{Output of a simple cell}

We obtain the output response of a simple cell (which is located at the point $q=(q_1,q_2)\in M\simeq \R^2$ and sensitive to the feature values $z=(\theta, \phi, \omega)$) to a generic image $I: M\rightarrow \R$ as a convolution with Gabor filter banks:
\begin{equation}\label{eq:outputExpressionExtended}
O^{I}(q,z)=\INT_{M} I(x,y)\Psi_{(q,z)}(x,y,s)\,dx\,dy.
\end{equation}
We apply the convolution for all feature values $z$ at every point $q$ in order to obtain the output responses of all receptive profiles in the set of receptive profiles. It is equivalent to applying a multi-frequency Gabor transform on the given two dimensional image. Since we use all frequency components of the transform, we can employ the exact inverse Gabor transform in order to obtain the initial image:
\begin{equation}\label{eq:inverseGaborTransformExpression}
I(q)=\INT_{\mathcal{M}} O^I(x,y, z)\bar{\Psi}_{(x,y,z)}(q,s)\,dx\,dy\,dz,
\end{equation}
with $\bar{\Psi}$ denoting the complex conjugate. We will call the output response \emph{lifted image} and the Gabor transform \emph{lifting}.

We remark here that we consider the whole complex structure of the result of the convolution \eqref{eq:outputExpressionExtended} as the output response of a simple cell. It is different from the cases of the previous visual cortex models which were choosing either real or imaginary part of the output responses obtained as the result of the convolution with corresponding Gabor filters (see for example \cite{citti2006cortical}, \cite{sarti2008symplectic}, \cite{sarti2009functional}). In other words they were not taking into account the half of the information which they obtained from an image. Furthermore, inverse Gabor transform was not possible in the previous models of visual cortex given in \cite{citti2006cortical}, \cite{sarti2008symplectic}, \cite{sarti2009functional} since in those models a single frequency Gabor transform was employed to obtain the output responses.

\subsection{Horizontal vector fields and connectivity}
Horizontal vector fields are defined as the elements of
\begin{equation}
\operatorname{ker}\Theta=\{X \in T\mathcal{M}:\; \Theta(X)=0\},
\end{equation}
where $T\mathcal{M}$ denotes the tangent bundle of the 5-dimensional manifold $\mathcal{M}$. They are induced naturally by the 1-form $\Theta$ given in \eqref{eq:one_form_theta}. The horizontal vector fields are found explicitly as
\begin{align}\label{eq:horizontalLIVFsExtended}
\begin{split}
 X_1 & =\cos\theta\,\partial_{x}+\sin\theta\,\partial_{y},\quad X_2  =\partial_{\theta},\\
 X_3 & =-\sin\theta\,\partial_x+\cos\theta\,\partial_y+\omg\,\partial_s,\quad  X_4  =\partial_{\omg}.
\end{split}
\end{align}
The corresponding horizontal distribution is therefore as follows:
\begin{equation}\label{eq:extendedHorizontalTangentSpace}
\mathcal{D}^{\mathcal{M}}=\operatorname{span}(X_1,X_2,X_3,X_4).
\end{equation}

All non-zero commutators related to the horizontal vector fields given in \eqref{eq:horizontalLIVFsExtended} follow as
\begin{align}
\begin{split}
[X_1, X_2]= & \sin\theta\,\partial_x-\cos\theta\,\partial_y,\\
[X_2, X_3]= & -\cos\theta\,\partial_sx-\sin\theta\,\partial_y,\\
[X_3, X_4]= & -\partial_s.
\end{split}
\end{align}

Note that the horizontal vector fields are bracket generating since 
\begin{equation}\label{eq:bracketGeneratingProperty}
T_\alpha\mathcal{M}=\operatorname{span}(X_1, X_2, X_3, X_4, [X_1, X_2])(\alpha),
\end{equation}
for all $\alpha\in \mathcal{M}$, where $T_\alpha\mathcal{M}$ denotes the tangent space of $\mathcal{M}$ at $\alpha$. Obviously \eqref{eq:bracketGeneratingProperty} shows that the horizontal vector fields fulfill the H\"{o}rmander condition \cite{hormander1967hypoelliptic}, and consequently they provide the connectivity of any two points on $\mathcal{M}$ through the horizontal integral curves defined on $\mathcal{M}$ due to the Chow's theorem \cite{chow2002systeme}. This connectivitiy property is particularly important since it guarantees that any two points in V1 can be connected via the horizontal integral curves, which are the models of the neural connectivity patterns in V1.

\subsection{Functional architecture of the visual cortex}
\subsubsection{The architecture as a Lie group}
Receptive profiles evoke a group structure at each frequency $\omg\in \R^+$. We can describe the group structure underlying the set of receptive profiles by using the transformation law given in \eqref{eq:coordinateTransformLocalGlobal}. 

First we notice that the elements $(q,\theta,\phi)$ induce the group given by
\begin{equation}
G_{\omg}\simeq \{A_{(q,\theta,\phi)}:\; (q,\theta,\phi)\in M\times S^1\times S^1\},
\end{equation}
which is indeed a Lie group associated to a fixed frequency $\omg$.

Then we write the group multiplication law for two elements
\begin{equation}
 g=(q^g,\theta_1,\phi_1),\quad h=(q^h, \theta_2,\phi_2),\quad g,h\in G_{\omg},
 \end{equation} 
as
 \begin{equation}
 g h=\big(\begin{pmatrix}
 q^g_1 \\ q^g_2
\end{pmatrix}+R_{\theta_1+\theta_2}\begin{pmatrix}
q^h_1 \\ q^h_2
\end{pmatrix},\; \theta_1+\theta_2,\; \phi_1+\phi_2 \big),
 \end{equation}
by using \eqref{eq:coordinateTransformLocalGlobal}.

The differential $L_{g^{\ast}}$ of the left-translation
\begin{align}
\begin{split}
L_g:  \;G_{\omg} & \rightarrow G_{\omg} \\
 h  & \mapsto g h ,
\end{split}
\end{align}
is  given by
\begin{equation}
L_{g^{\ast}}=\begin{pmatrix}
\cos(\theta) & 0 & -\sin(\theta) & 0\\
\sin(\theta) & 0 & \cos(\theta) & 0 \\
0 & 1 & 0 & 0\\
0 & 0 & \omg & 0
\end{pmatrix}.
\end{equation}

The vector fields $X_1$, $X_2$ and $X_3$ are bracket generating due to that
\begin{equation}
\operatorname{span}(X_1, X_2, X_3, [X_1, X_2] )(g)=T_gG_{\omg},
\end{equation}
for every $g\in G_{\omg}$. Hence $X_1$, $X_2$ and $X_3$ generate the Lie algebra corresponding to $G_{\omg}$.

\subsubsection{The architecture as a sub-Riemannian structure}
The functional geometry is associated to a sub-Riemannian structure at each frequency $\omg$. We denote by $G_{\omg}$ a submanifold of $\mathcal{M}$ with points $h=(q,\theta,\phi,\omg)=(q,z)$ restricted to a fixed $\omg$. In this case the horizontal distribution is found by
\begin{equation}
 \mathcal{D}^{G_{\omg}}=\operatorname{span}( X_1, X_2, X_3 ) .
\end{equation} 
 
Furthermore the induced metric $(g_{ij})^{G_{\omg}}_h: \mathcal{D}^{G_{\omg}}\times  \mathcal{D}^{G_{\omg}}\rightarrow \R$ is defined on $\mathcal{D}^{G_{\omg}}$ and at every point $h\in G_{\omega}$ 
will make $X_1, X_2, X_3 $ orthonormal. 

Finally the associated sub-Riemannian structure to the frequency $\omg$ is written as the following triple:
\begin{equation}
(G_{\omg}, \mathcal{D}^{G_{\omg}}, (g_{ij})_h^{G_{\omg}}).
\end{equation}

\section{Horizontal integral curves}\label{sec:Connectivity_in_the_extended_phase_space}

The lifting mechanism leaves each lifted point isolated from each other since there is no connection between the lifted points. Horizontal vector fields endow the model with an integration mechanism which provides an integrated form of the local feature vectors obtained from the lifted image at each point on $\mathcal{M}$.

Once a simple cell is stimulated, its activation propagates between the simple cells
along certain patterns which can be considered as the integrated forms of the local feature vectors. This propagation machinery is closely related to the association fields \cite{field1993contour}, which are the neural connectivity patterns between the simple cells residing in different hypercolumns (long range horizontal connections) within V1. The association fields coincide with the anisotropic layout of the long range horizontal connections at the psychophysical level. In the classical framework of \cite{citti2006cortical}, those association fields were modeled as the horizontal integral curves of $\se$. We follow a similar approach and propose to model the association fields in our model framework as the horizontal integral curves associated to the 5-dimensional sub-Riemannian geometry of $\mathcal{M}$. We conjecture that those horizontal integral curves coincide with the long range horizontal connections between orientation, frequency and phase selective simple cells in V1.

We denote a time interval by $\mathcal{I}=[0,T]$ with $0<T<\infty$ and then consider a horizontal integral curve $(q_1,q_2,\theta,\omg,\phi)=\gamma:\mathcal{I}\rightarrow\mathcal{M}$ associated to the horizontal vector fields given in \eqref{eq:horizontalLIVFsExtended} and starting from an initial point $\hat{\alpha}=(\hat{q}_1,\hat{q}_2,\hat{\theta},\hat{\omg},\hat{\phi})$. Let us denote the velocity of $\gamma$ by $\gamma^{\prime}$. At each time $t\in \mathcal{I}$ the velocity is a vector $\gamma^{\prime}(t)\in \operatorname{span}(X_1,X_2,X_3,X_4)\big(\gamma(t) \big)$ at $\gamma(t)=(q_1(t),q_2(t),\theta(t),\omg(t),\phi(t))\in\mathcal{M}$. In order to compute the horizontal integral curves, we first consider the vector field $\gamma^{\prime}$ which is given by
\begin{align}
\gamma^{\prime}(t)=X(\gamma(t))=
(c_1 X_1+ c_2 X_2+c_3 X_3+c_4 X_4)(\gamma(t)),\quad t\in \mathcal{I},
\label{eq:referredAtFanPlot}
\end{align} 
with coefficients $c_i$ (which are not necessarily constants) where $i\in\{1,2,3,4 \}$. Then we can write each component of $\gamma^{\prime}(t)$ as follows:
\begin{align}\label{eq:odeSys1}
\begin{split}
q^{\prime}_1(t) & =c_1\cos(\theta(t))-c_3\sin(\theta(t)),\\
q^{\prime}_2(t) & =c_1\sin(\theta(t))+c_3\cos(\theta(t)),\\
\theta^{\prime}(t) & = c_2,\\
\omg^{\prime} (t) & = c_4,\\
\phi^{\prime}(t) & = c_3\,\omg(t).
\end{split}
\end{align}
In the case of that the coefficients $c_i$ are real constants and $c_2\neq 0$, we solve the system of the ordinary differential equations (given in \eqref{eq:odeSys1}) of $t$ with the initial condition $\hat{\alpha}$ and find the solution as follows:
\begin{align}\label{eq:odeSoln1}
\begin{split}
q_1(t) & =\hat{q}_1+\frac{1}{c_2}\big(-c_3\cos(\hat{\theta})+c_3\cos(c_2 t+\hat{\theta})-c_1\sin(\hat{\theta})+c_1\sin(c_2 t+\hat{\theta})      \big),\\
q_2(t) & =\hat{q}_2+\frac{1}{c_2}\big(c_1\cos(\hat{\theta})-c_1\cos(c_2 t+\hat{\theta})-c_3\sin(\hat{\theta})+c_3\sin(c_2 t+\hat{\theta})\big),\\
\theta(t) & = c_2 t+\hat{\theta},\\
\omg(t) & = c_4 t+\hat{\omg},\\
\phi(t) & = \frac{1}{2}\big(c_3 c_4 t^2+2 t c_3 \hat{\omg}+2\hat{\phi}  \big).
\end{split}
\end{align}

If $c_2=0$ then the solution becomes
\begin{align}\label{eq:odeSoln2}
\begin{split}
q_1(t) & =\hat{q}_1+t\big(c_1\cos(\hat{\theta})-c_3\sin(\hat{\theta})\big),\\
q_2(t) & =\hat{q}_2+t\big( c_3\cos(\hat{\theta})+c_1\sin(\hat{\theta})\big),\\
\theta(t) & =\hat{\theta},\\
\omg(t) & =c_4 t+\hat{\omg},\\
\phi(t) & =\frac{1}{2}(c_3c_4 t^2+2 t\, c_3\,\hat{\omg}+2\hat{\phi}).
\end{split} 
\end{align}

\begin{figure}[htp]
\centerline{\includegraphics[scale=0.6,trim={0cm 0 0 0},clip]{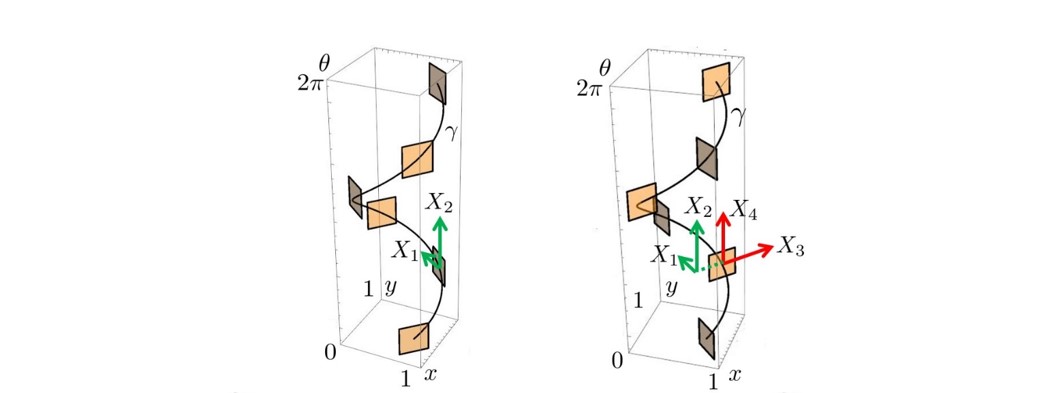}}
\caption{ An integral curve along the vector field $X_1+X_2$. It represents an orientation fiber once $\omg$ and $\phi$ are fixed. The tangent planes spanned by $X_1,$ $X_2$ (left) and $X_3,$ $X_4$ (right) are shown at six points on the curve.}
\label{fig:orthTan2}
\end{figure}

\begin{figure}[htp]
\centerline{\includegraphics[scale=0.325,trim={0cm 0 0 0},clip]{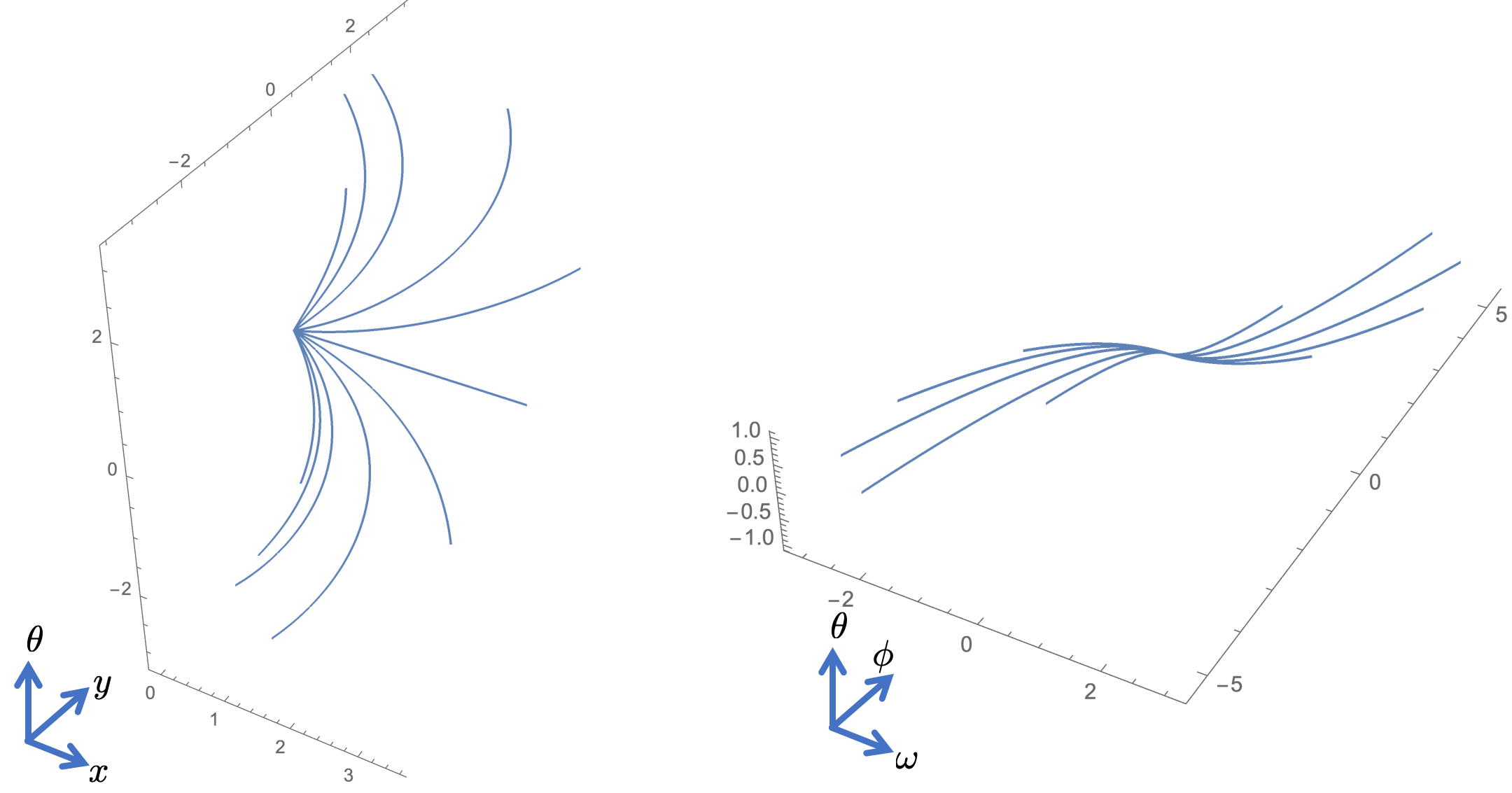}}
\caption{Integral curve fans corresponding to $X_1+c_2 X_2$ (left) and $X_3+c_4 X_4$ (right) where $c_2$ and $c_4$ are varied, respectively.}
\label{fig:integralFans}
\end{figure}

Note that \eqref{eq:odeSoln1} and \eqref{eq:odeSoln2} describe the whole family of the horizontal integral curves described by the horizontal distribution 
$$\mathcal{D}^{\mathcal{M}}=\displaystyle\bigcup_{\omg\in\R^+}\mathcal{D}^{G_{\omg}}=\operatorname{span}( X_1, X_2, X_3, X_4).$$
We are interested rather in two specific sub-families corresponding to the horizontal vector fields which reside in either one of the two orthogonal $\mathcal{D}_\alpha^{\mathcal{M}}$ subspaces which are defined at every point $\alpha=(q,\theta,\omg,\phi)\in \mathcal{M}$ as
\begin{equation}
S_1\mathcal{D}^{\mathcal{M}}_\alpha=\operatorname{span}(X_1,X_2)(\alpha),\quad S_2\mathcal{D}^{\mathcal{M}}_\alpha=\operatorname{span}(X_3, X_4)(\alpha),
\end{equation}
satisfying
\begin{equation}
\mathcal{D}^{\mathcal{M}}_\alpha=S_1\mathcal{D}^{\mathcal{M}}_\alpha\oplus S_2\mathcal{D}^{\mathcal{M}}_\alpha.
\end{equation}
Figure \ref{fig:orthTan2} gives an illustration of the orthogonal layout of $S_1\mathcal{D}^{\mathcal{M}}_\alpha$ and $S_2\mathcal{D}^{\mathcal{M}}_\alpha $ at points $\alpha$ on an orientation fiber, i.e., on a horizontal integral curve along $X_1+X_2$ corresponding to some fixed $\omg$ and $\phi$. See also Figure \ref{fig:integralFans}, where the integral curves along the vector fields $X_1+c_2 X_2$ and $X_3+c_4 X_4$ with varied $c_2$ and $c_4$ values, respectively, are presented. 

We remark here that $S_1\mathcal{D}^{\mathcal{M}}_\alpha $ is the horizontal tangent space $T_{(q,\theta)}\se$ of $\se$ at the point $\alpha$ once frequency $\omg$ and phase $\phi$ are fixed. In other words at each point $\alpha=(q,\theta,\omg,\phi)$ with $\omg$ and $\phi$ fixed on $\mathcal{M}$, one finds the sub-manifold $\se$ which is the classical sub-Riemannian geometry corresponding to the model given in \cite{citti2006cortical}. This property allows the simple cell activity to be propagated in each subspace corresponding to a frequency-phase pair separately and it will be important for image enchancement applications employing our model framework.

\FloatBarrier

\section{Enhancement}\label{sec:EnhancementSection}

Image enhancement refers to smoothing a given input image, reducing the noise and at the same time preserving the geometric structures (edges, corners, textures and so on). We perform our image enhancement procedure on the output responses instead of on the input image. 
Since the output responses encode the local feature values of orientation, frequency and phase, this allows us to exploit the additional information obtained from those features.  Our enhancement procedure is based on an iterative Laplace-Beltrami procedure on the simple cell output responses in the 5-dimensional sub-Riemannian geometry $\mathcal{M}$ and it results in a mean curvature flow in the geometry.

\subsection{Laplace Beltrami procedure}
Anisotropic metric on the space $\mathcal{M}$ of simple cell output responses defines the sub-Riemannian Laplacian in the sub-Riemannian space generated by the simple cells: 
\begin{equation}\label{eq:HorizontalLaplacianWithCoeffs}
\Delta_0 u=\SUM_{i=1}^4 c_i X_{i}X_i u,
\end{equation}
where coefficients $c_i$ are non-negative constants representing the weights of the second order horizontal vector fields which are given in \eqref{eq:horizontalLIVFsExtended}.
The weights are used to adjust the operator to the sub-Riemannian homogeneity of $\mathcal{M}$. They are particularly important in the discrete case, where different dimensions of the space need not necessarily be sampled in the same way.

It has been proved by Franceschiello et. al. in \cite{franceschiello2018neuromathematical} that the output induces a metric on the space of the model geometry proposed in \cite{citti2006cortical} and the metric elicits certain visual illusions. In the article of Franceschiello et. al. \cite{franceschiello2018neuromathematical} a simplified diagonal metric was used. On the other hand, following the approach of Kimmel et. al. \cite{kimmel2000geometry}, \cite{kimmel2000images}, we choose the metric induced by the output $O^I(q, z)$ on $\mathcal{M}$ and use a simplified version of this metric for the applications. 

The metric $(g_{ij})$ induced by the output responses is defined as follows:
\begin{definition}\label{def:srKimmelMetricDefinition}
\begin{equation}
(g_{ij})=\begin{pmatrix}1+c_1(X_1u)^2 & \sqrt{c_1c_2}X_1uX_2 u & \sqrt{c_1c_3}X_1uX_3 u & \sqrt{c_1c_4}X_1uX_4 u \\
\sqrt{c_1c_2} X_2u X_1 u & 1+ c_2 (X_2u)^2 & \sqrt{c_2c_3} X_2uX_3 u & \sqrt{c_2c_4} X_2uX_4 u \\
\sqrt{c_1c_3} X_3u X_1 u & \sqrt{c_2c_3} X_3uX_2 u & 1+ c_3 (X_3 u)^2 & \sqrt{c_3c_4} X_3uX_4 u \\ 
\sqrt{c_1c_4} X_4uX_1 u  & \sqrt{c_2c_4} X_4uX_2 u & \sqrt{c_3c_4}X_4uX_3 u & 1+c_4 (X_4 u)^2
\end{pmatrix},
\end{equation}
with constants $c_1,c_2,c_3,c_4\geq 0$.
\end{definition}
We denote the inverse metric by $(g^{ij})$ and its elements by $g^{ij}$.

Mean curvature flow provides an adapted enhancement to the surface underlying the image function $I$ since the flow is restricted to the evolving level sets of the image. Laplace-Beltrami operator is written as: 
\begin{equation}\label{eq:laplaceBeltramiOperator}
L u=\SUM_{i,j=1}^4\displaystyle\frac{1}{\sqrt{\operatorname{det}(g_{ij})}}X_i\big(\sqrt{\operatorname{det}(g_{ij})}g^{ij}X_j u\big ),
\end{equation}
where $\operatorname{det}(g_{ij})$ is the determinant of the induced metric. Laplace-Beltrami operator can be considered as the linearization of the motion by curvature explained in \cite{baspinar2016uniqueness}. For practial reasons, we will use a Laplace-Beltrami process with the operator given in \eqref{eq:laplaceBeltramiOperator} associated to a reduced version of the metric provided in Definition \ref{def:srKimmelMetricDefinition}.

The evolution equation for the enhancement via sub-Riemannian Laplace-Beltrami procedure is written as:
\begin{equation}\label{eq:generalEvolutionEqn}
\begin{cases}
\partial_t u=L\,u\\
u_{|t=0}=O^{I}(q, p),
\end{cases}
\end{equation}
for all $(q,p)\in\mathcal{M}$ and $0<t\leq T$.

\subsubsection{Reduced equation}
\label{sec:reducedEquation}
It is possible to perform the Laplace-Beltrami procedure in each frequency and phase sub-space separately in a reduced framework. In that case we choose $c_1,c_2> 0$ and $c_3=c_4=0$. In this way we apply the evolution equation on surfaces in each frequency and phase sub-space, i.e., on each $\se_{(\omg,\phi)}$ manifold, which is the submanifold with elements $(q,\theta)$ representing the points $(q,\theta,\omg,\phi)\in \mathcal{M}$ with fixed $\omg$ and $\phi$. In this framework the metric $(g_{ij})$ boils down to
\begin{equation}\label{eq:metricForApproximation}
(g_{ij})=\begin{pmatrix} 1+c_1 (X_1 u)^2 & \sqrt{c_1c_2}X_1uX_2u \\
\sqrt{c_1c_2}X_2uX_1 u &  1+c_2 (X_2 u)^2
\end{pmatrix}.
\end{equation}
We choose $c_1$ and $c_2$ suitably by regarding the fixed $\omega$ values.

The motivation for choosing $c_3=0$ is that we would like to avoid excessive diffusion in the direction of the vector field $X_3$. We already have sufficient diffusion in this direction due to the commutator $[X_1, X_2]$. Direct application of $X_3$ introduces additional diffusion in ortogonal directions to the object boundaries, which is not desired since it might destroy object boundaries and contour structures in the input image. Furthermore, the use of the reduced version lowers the computational load since now multiple Laplace-Beltrami procedures are applied in 3-dimensional sub-Riemannian geometry $\se_{(\omega, \phi)}$ at each frequency $\omega$ instead of in the 5-dimensional sub-Riemannian geometry $\mathcal{M}$.

We remark that the vector field $X_3$ does not perform information flow only in the orthogonal direction
\begin{equation}
-\sin\theta \partial_x+\cos\theta\partial_y,
\end{equation}
to the boundaries but also in the direction of phase. However the elimination of $X_3$ from the Laplace-Beltrami procedure must be accordingly taken into account also in the metric given in Definition \ref{def:srKimmelMetricDefinition} in order to provide the coherency between the Laplace-Beltrami operator and the employed metric. This is the reason for that we fix $c_3=0$ in the reduced version of the metric given in \eqref{eq:metricForApproximation}.


We also choose $c_4=0$. Indeed we assume that no information flow takes place  along the vector field $\partial_s$, i.e., in the phase direction. 
We notice that the Gabor transform produces a rotated version of the image $I$ by the angle $\phi$ for each phase (see \cite{duits2013evolution} for more details). 
Hence the Laplace-Beltrami procedure is applied on the rotated versions of the same initial image and the result is the same but only rotated for each phase value $\phi$.

Although in the present study we will not provide any results related to image inpainting task of the Laplace-Beltrami procedure, we would like to mention a few related points. The use of the vector field $X_3$ becomes important in texture image inpainting. In that case, on the contrary to the enhancement, we would like to have information flow in orthogonal directions to the object boundaries and reduce the flow along the boundaries. In that case, since also the spatial frequency of the texture patterns have a great importance, we would like to keep the track of the frequency as well as the phase of the evolving output responses, and we would need to fix $c_1$ and $c_2$ to zero instead of $c_3$ and $c_4$ in that case.

\subsection{Implementation of the algorithm}\label{sec:Algorithms}
\subsubsection{The algorithm}
\label{sec:recipe_algorithm}
We present the steps of our algorithm based on \eqref{eq:generalEvolutionEqn} by starting from the initial image function $I:\R^2\simeq M\rightarrow \R$ at $q\in M$.

\begin{enumerate}
\item Lift the image $I(q)$ to $O^{I}(q, p)$ by using \eqref{eq:outputExpressionExtended}. Choose this 
output as the initial value $u_{\vert t=0}$ of the solution to \eqref{eq:generalEvolutionEqn} at time $t=0$. 
 
\item Denote the discrete step in time by $\Delta t$. At the $k^{\text{th}}$ iteration (i.e., $t=k\Delta t$) compute the result of the discretized version $\bar{L}$ (of the operator $L$) applied on the current value of $u$ at time instant $t$ as $\bar{L} u(t)$ and update the solution and the value of $u(t)$ by using \eqref{eq:generalEvolutionEqn} as follows:  
$$u(t+\Delta t)=u(t) +\Delta t \bar{L}u(t).$$
\item Repeat step 2 until the final time $T=(\text{number of  iterations})\times\Delta t$ is achieved.
\item Apply the inverse Gabor transform given by \eqref{eq:inverseGaborTransformExpression} on $u(T)$.
\end{enumerate}

\subsubsection{Discrete simple cell output responses}
\label{sec:Discrete_Gabor_coeff}
We discretize the image function $I$ on a uniform spatial grid  as
\begin{equation}
I[i,j]=I(i\Delta x, j\Delta y),
\end{equation}
with $i,j\in \{1,2,\dots,N\}$ ($N$ is the number of samples in spatial dimensions) and $\Delta x,\Delta y\in \R^+$ denoting the pixel width (In general we use square images as input image and we fix $\Delta x=\Delta y=1$ in terms of pixel unit). Furthermore the discretized simple cell response $O^I(q_{1,i},q_{2,j},\theta_k,\omg_l,\phi_m)$ of $I[i,j]$ on uniform orientation, frequency and phase grids with points $\theta_k=k\Delta\theta$, $\omg_l=l\Delta\omg$ and $\phi_m=m\Delta s$ ($k\in \{1,2,\dots, K\}$, $l\in \{1,2,\dots,L\}$, $m\in \{1,2,\dots, M\}$ (where we denote the number of samples in the orientation dimension by $K$, in the frequency dimension by $L$ and in the phase dimension by $M$, and the distances between adjacent samples in the orientation dimension by $\Delta\theta$, in the frequency dimensions by $\Delta \omg$ and in the phase dimension by $\Delta s$) is denoted by
\begin{equation}
O^I[i,j,k,l,m]=O^I(q_{1,i},q_{2,j},\theta_k,\omg_l,\phi_m),
\end{equation}
where $q_{1,i}=i\Delta x$ and $q_{2,j}=j\Delta y$.

In this case the discrete version of the Gabor function given by \eqref{eq:generalGaborFromMother} is written as:
\begin{equation}
\Psi_{[i,j,k,l,m]}[\tilde{i},\tilde{j},\tilde{n}]=\Psi_{(q_{1,i},\,q_{2,j},\,\theta_k,\,\omg_l,\,\phi_m)}(\tilde{x}_{\tilde{i}},\tilde{y}_{\tilde{j}},\tilde{s}_{\tilde{n}}),
\end{equation}
where $\tilde{i},\tilde{j}\in\{1,2,\dots,\tilde{N}\}$, $\tilde{k}\in\{1,2,\dots,\tilde{K}\}$, $\tilde{n}\in \{1,2,\dots,\tilde{M}\}$.
Then we fix $s_{\tilde{n}}=0$ (i.e., $\tilde{n}=0$) in the reduced framework (which was explained in Section \ref{sec:reducedEquation}) and write the discrete cell response obtained from the image $I[i,j]$ via the discrete Gabor transform as:
\begin{equation}
O^I[i,j,k,l,m]=\SUM_{\tilde{i},\tilde{j}}\Psi^l_{[i,j,k,m]}[\tilde{i},\tilde{j},0]\,I[\tilde{i},\tilde{j}].
\end{equation}
The time correspondence in the discrete case is represented by the time index $h_p$ where the time interval is discretized by $P\in\mathbb{N}^+$ samples and $h_p$ represents the time instant $h_p=p\Delta t$ with $\Delta t$ satisfying $T=P\Delta t$ and $p\in\{1,2,\dots,P\}$. In this case the discretized Gabor coefficient is written as
\begin{equation}
O^{I,h_p}[i,j,k,l,m]=O^{I,h_p}(q_{1,i},q_{2,j},\theta_k,\omg_l,\phi_m)=u(t+p\Delta t).
\end{equation}

\subsubsection{Explicit scheme with finite differences}
 Here we provide the discrete scheme related to the numerical approximation of the algorithm. We propose an explicit finite difference scheme in order to iterate the evolution equation given in \eqref{eq:generalEvolutionEqn}. The reason for choosing explicit scheme is that implicit scheme requires large memory in our 4-dimensional (reduced) anisotropic framework.

We obtain the explicit scheme first by writing \eqref{eq:generalEvolutionEqn} in terms of the horizontal vector fields $X_1$, $X_2$, $X_3$ and $X_4$ given in \eqref{eq:horizontalLIVFsExtended}. Then following Unser \cite{unser1999} and Franken \cite{franken2008enhancement}, we implement the horizontal vector fields by using central finite differences which are interpolated by B-splines on a uniform spatial sample grid. Note that B-spline interpolation is required since not all horizontal vectors are aligned with the spatial sample grid.

The interpolation is achieved by determining the coefficients $b(i,j)$
\begin{equation}
s(x,y)=\SUM_{i,j\in Z}b(i,j)\rho(x-i,y-j),
\end{equation}
in such a way that the spline polynomial $s(x,y)$ with the B-spline basis functions $\rho(x-i, y-j)$ coincides with the horizontal derivatives of the output $O^I$ at the grid points. For example, in the case of the first horizontal derivative $X_1 O^{I}$, the condition $s(i\Delta x, j\Delta y)=X_1O^{I}[i,j,k,l,m]$ must hold if we consider a discrete output as explained in Section \ref{sec:Discrete_Gabor_coeff}. We refer to the explanations of Unser \cite{unser1999} for details.

\begin{figure}[htp]
\centerline{\includegraphics[scale=0.370,trim={0cm 0cm 0cm 0cm},clip]{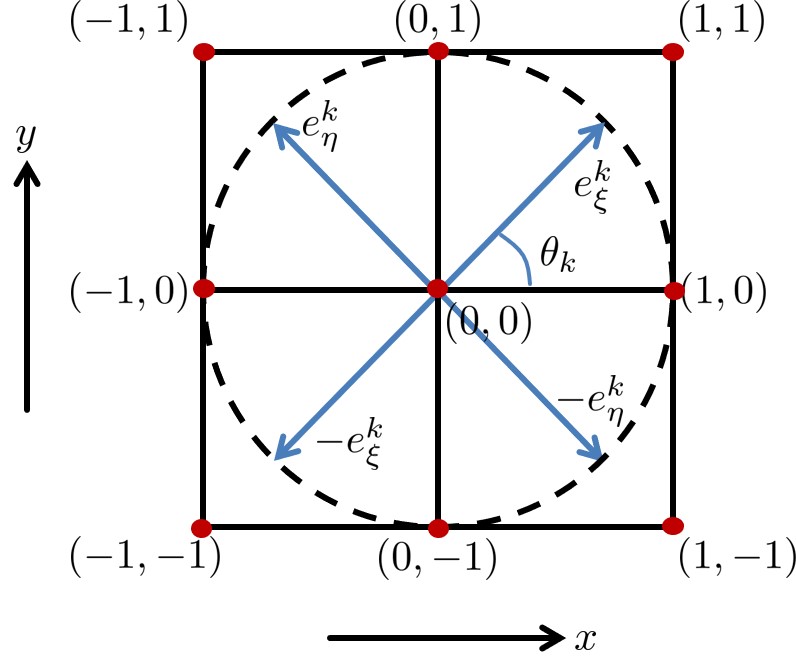}}
\caption{(Adapted from Franken \cite{franken2007nonlinear}) Illustration of the vectors $e^k_{\xi}$ and $e_{\eta}^k$ at $(0,0)$ with $\Delta x=\Delta_y=1$.}
\label{fig:finiteDifferenceScheme2}
\end{figure}

We fix $\Delta x=\Delta y=1$ and define
\begin{align}
\begin{split}
e^k_{\xi}:= & (\Delta x\cos(\theta_k),\Delta y\sin(\theta_k)),\\
e^k_{\eta}:= & (-\Delta x\sin(\theta_k),\Delta y\cos(\theta_k)).
\end{split}
\end{align}
See Figure \ref{fig:finiteDifferenceScheme2} for an illustration of those vectors. We write the central finite differences of the first order horizontal derivatives as
\begin{align}
\begin{split}
X_1 O^{I,h_p}[i,j,k,l,m]\approx & \frac{1}{2\Delta x}(O^{I,h_p}(q+e^k_{\xi},\theta_k,\omg_l,\phi_m) -O^{I,h_p}(q-e^k_{\xi},\theta_k,\omg_l,\phi_m)),\\
X_2 O^{I,h_p}[i,j,k,l,m]\approx & \frac{1}{2\Delta \theta}(O^{I,h_p}(q,\theta_{k+1},\omg_l,\phi_m) -O^{I,h_p}(q,\theta_{k-1},\omg_l,\phi_m)),
\end{split}
\end{align}
and of the second order horizontal derivatives which we use as
\begin{align}
\begin{split}
X_1X_1O^{I,h_p}[i,j,k,l,m]\approx \frac{1}{(\Delta x)^2}  &  \Big(O^{I,h_p}(q+e^k_{\xi},\theta_k,\omg_l,\phi_m)-2O^{I,h_p}(q,\theta_k,\omg_l,\phi_m)\\ & +O^{I,h_p}(q-e^k_{\xi},\theta_k,\omg_l,\phi_m)\Big),\\
X_2X_2 O^{I,h_p}[i,j,k,l,m]\approx \frac{1}{(\Delta \theta)^2}  &  \Big(O^{I,h_p}(q,\theta_{k+1},\omg_l,\phi_m)-2 O^{I,h_p}(q,\theta_{k},\omg_l,\phi_m)\\   & +O^{I,h_p}(q,\theta_{k-1},\omg_l,\phi_m)\Big).
\end{split}
\end{align}
Then the numerical iteration (discretized from step 2 of the algorithm provided in Section \ref{sec:recipe_algorithm}) with a time step $\Delta t>0$ is written as follows:
\begin{equation}\label{eq:discreteGeneralExpression}
\begin{split}
O^{I,p+1}[i,j,k,l,m]= & O^{I,h_{p+1}}(q_{i,1},q_{j,2},\theta_{k},\omg_l,\phi_m) \\
= & O^{I,h_p}{(q_{i,1},q_{j,2},\theta_{k},\omg_l,\phi_m)}+\Delta t \,\bar{L} O^{I,h_p}(q_{i,1},q_{j,2},\theta_{k},\omg_l,\phi_m),
\end{split}
\end{equation}
where $\bar{L}$ represents the discretized version of $L$ given in \eqref{eq:laplaceBeltramiOperator} (with coefficients $c=\{c_1>0,c_2>0,c_3=0,c_4=0\}$) in terms of the central finite differences.

\subsubsection{Stability analysis}

We must consider two points for the stability of our finite discrete scheme:
\begin{enumerate}
\item Suitable choice of the time step size $\Delta t$,
\item Preserving the space homogeneity during the Laplace-Beltrami evolution.
\end{enumerate}

The stability analysis for the $\se$ case is explained in \cite{franken2008enhancement} and \cite{duits2010left} based on Gershgorin theory. We adapt this technique to our reduced framework and find the upper limit for the time step $\Delta t$ as:
\begin{equation}
\Delta t\leq\frac{2\left(\frac{s_{\theta}}{\beta}\right)^2}{4+4(1+\sqrt{2})\left(\frac{s_{\theta}}{\beta}\right)^2},
\end{equation}
where $s_{\theta}=\frac{2\pi}{K}$ is the sampling distance between adjacent orientation samples, $K\in \mathbb{N}^+$ denotes the number of the orientation samples and $\beta$ is the ratio between orientation and spatial samples. Parameter $\beta$ is either $1/8$ or $1/4$ in our experiments, yielding the condition $\Delta t\leq 0.17$ for stable processes for both $\beta$ values. We refer to \cite[Chapter 6]{franken2008enhancement} and \cite{creusen2011numerical} for details.

The second point is due to that we sample each dimension by using  a different number of samples. In order to perform sub-Riemannian diffusion by regarding the sample unit coherency one must choose the parameters $c_1$, $c_2$ of the operator $L$ in such a way that the space homogeneity of $\mathcal{M}$ is preserved.

\subsection{Experiments}
\label{sec:Experimental_results_image_processing}

\subsubsection{Gabor transform}

The delicate point related to the lifting and inversion process is that Gabor functions $\Psi_{(q,\theta,\omg,\phi)}(x,y,s)$ must be sampled (in orientation $\theta$, frequency $\omg$ and phase $\phi$ dimensions) in such a way that they cover all the spectral domain (that is, they must fulfill the Plancherel's formula \cite{plancherel1910contribution}). 

\begin{figure}[t]
\vspace{-1cm}
\centerline{\includegraphics[scale=0.370,trim={0cm 0cm 0cm 0cm},clip]{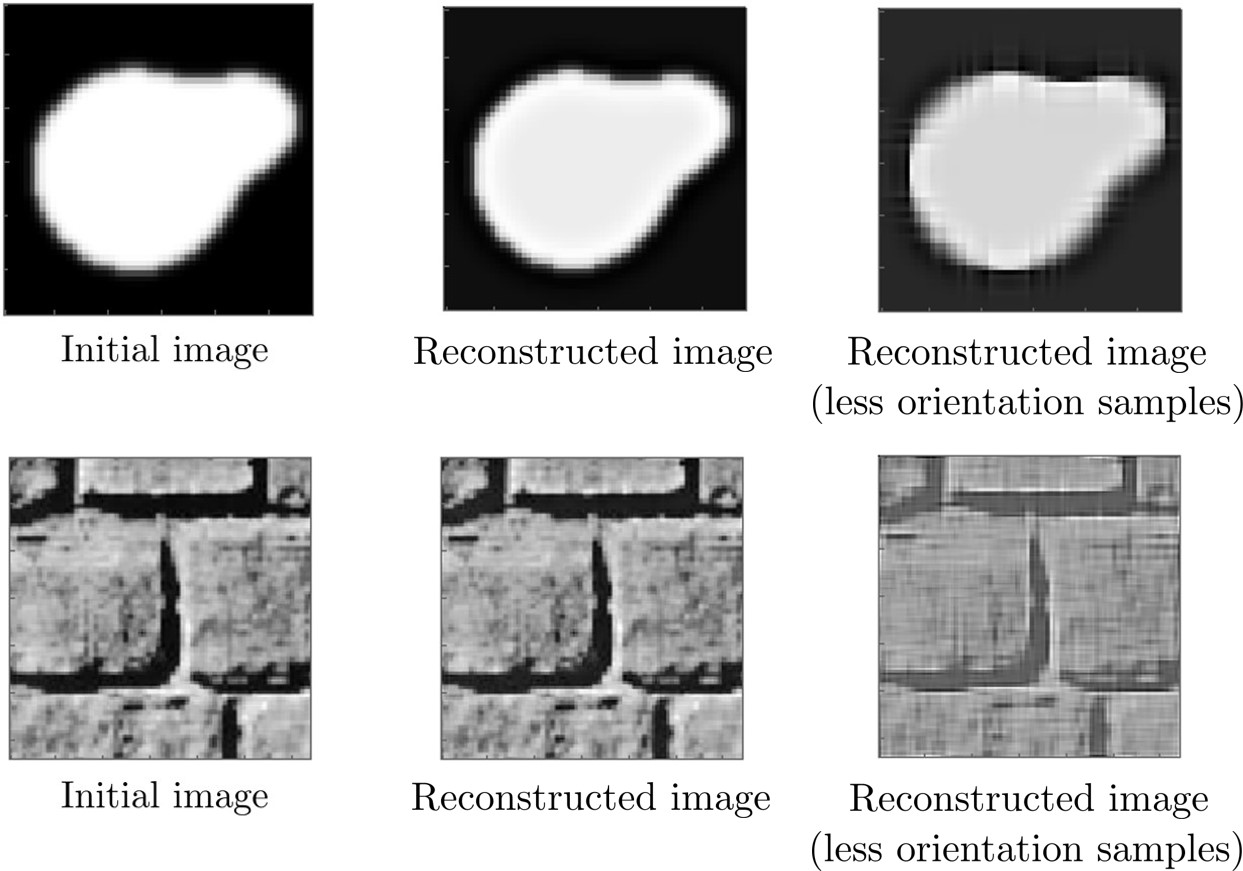}}
\caption{Examples of reconstructed images via transform and inverse transform procedure with Gabor functions, and the effect of number of orientation samples.}
\label{fig:gaborTransformInverse2}
\vspace{1cm}
\centerline{\includegraphics[scale=0.4,trim={0cm 0cm 0cm 0cm},clip]{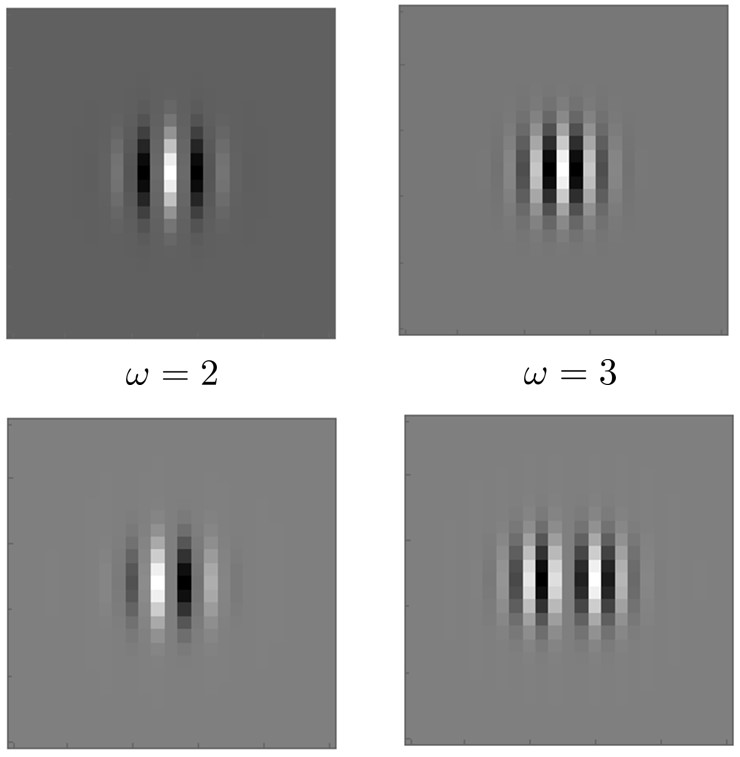}}
\caption{Examples of the Gabor filters used in the lifting procedure of Figure \ref{fig:gaborTransformInverse2}. Top: Even parts of the Gabor functions with frequencies $\omega=2,3$. Bottom: Odd parts of the same Gabor functions.}
\label{fig:gaborsUsed}
\end{figure}

We present some results of the Gabor transform-inverse transform procedure associated to our setting and the effects of number of samples in the orientation dimension in Figure \ref{fig:gaborTransformInverse2}. We use the Gabor filter banks obtained from \eqref{eq:extendedMotherGaborFcn} and \eqref{eq:generalGaborFromMother} with scale value of 2 pixels (total filter size is 24 pixels) in order to lift the test images (see Figure \ref{fig:gaborsUsed} for some examples of those Gabor functions). On the top row, we see the results related to an artificial $64\times 64$ test image (left), and at the bottom we see the results related to a real $64\times 64$ test image (left) taken from Kimmel et. al. \cite{kimmel2000images}  We see in the middle and right columns those two images now transformed and then inverse transformed with different number of orientation samples. We sample the space at frequencies  $\omg\in\{0.5, 1,\dots, 2.5, 2.75,\dots, 4.5,4.625,\dots, 6.5\}$, orientations $\theta\in\{\frac{2\pi}{32},\frac{4\pi}{32},\dots,\frac{62\pi}{32}\}$ (middle), $\theta\in \{0,\frac{2\pi}{8},\dots, \frac{14\pi}{8} \}$ (right) and phases $\phi\in \{0, \frac{\pi}{8},\dots, \frac{15\pi}{8}\}$. We observe that the decrease in the number of orientation samples reduce the quality of the tansformation procedure noticeably in both test images.

\FloatBarrier

\subsubsection{Enhancement}

The lifting procedure is performed by the Gabor filters of the type given by \eqref{eq:extendedMotherGaborFcn} and \eqref{eq:generalGaborFromMother} with $\text{scale}=2$ pixels (the filter size is $12\times \text{scale}=24$ pixels) and time step $\Delta t=0.1$ in the experiments.

\begin{figure}[htp]
\centerline{\includegraphics[scale=0.7,trim={0cm 0cm 0cm 0cm},clip]{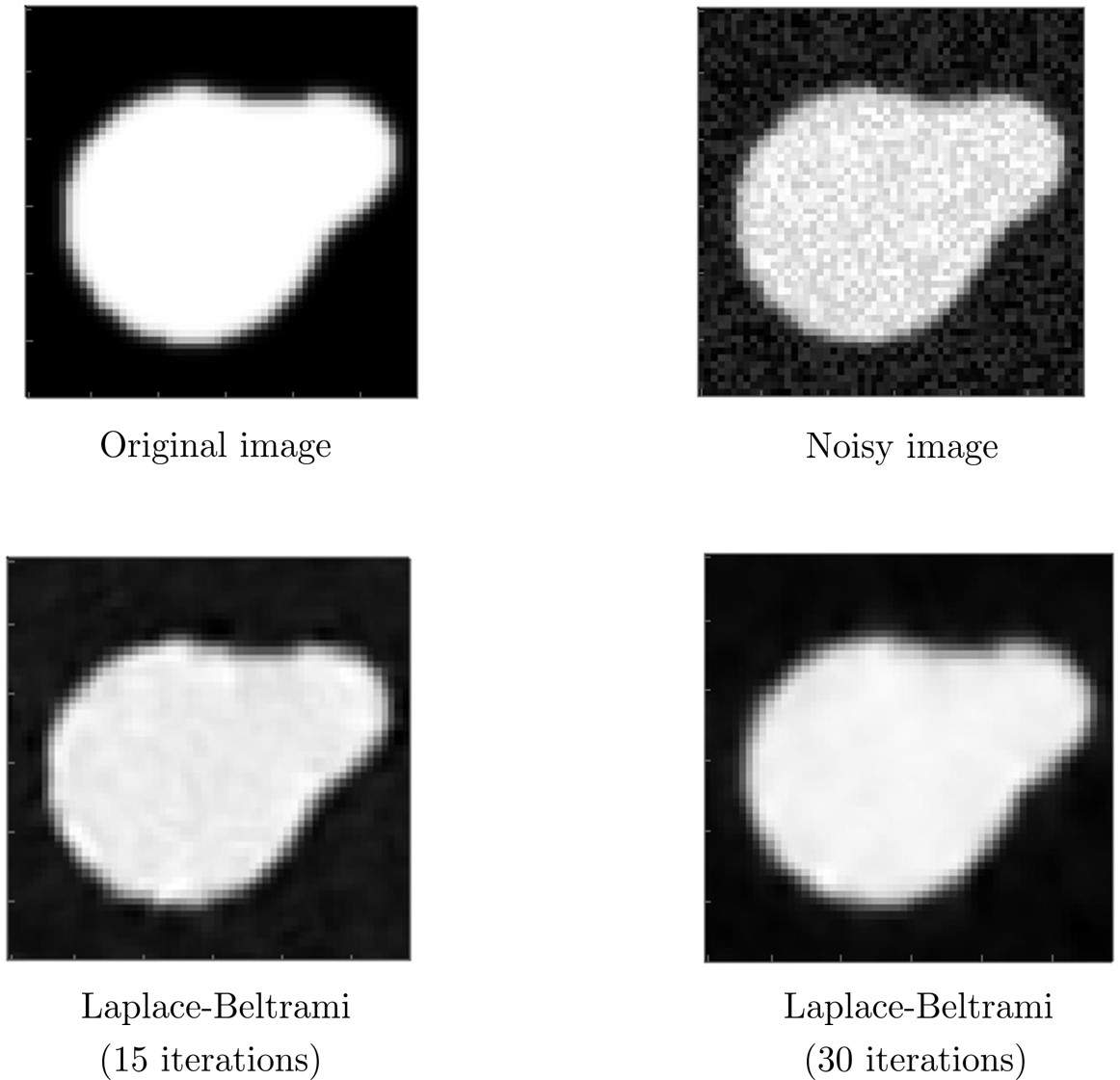}}
\caption{Top: The original $64\times 64$ image (left) and the noisy version (right). Bottom: The results of the Laplace-Beltrami procedure.}
\label{fig:comparisonDiffandLaplaceBeltrami2}
\end{figure}

In Figure \ref{fig:comparisonDiffandLaplaceBeltrami2}, we see the results by of the enhancement procedure applied on an artificially produced $64\times 64$ gray scale test image with white noise. The lifting is achieved with frequency samples $\omg\in\{0.5, 1,\dots, 2,2.25,\dots, 4.5 \}$, phase samples $\phi=\{ 0,\frac{\pi}{8},\dots,\frac{\pi}{2}\}$ and orientation samples $\theta\in\{0, \frac{2\pi}{16},\frac{4\pi}{16},\dots,\frac{30\pi}{16}\}$. Note that $\text{number of orientations}=16$, thus $\beta=\frac{\text{number of orientations}}{\text{image size}}=0.25$.
In order to fulfill physical unit coherency we choose $c_1=1$ and $c_2=\beta^2$. The experiments are done with 15 and 30 iterations.

\begin{figure}
\centerline{\includegraphics[scale=0.470,trim={0cm 0cm 0cm 0cm},clip]{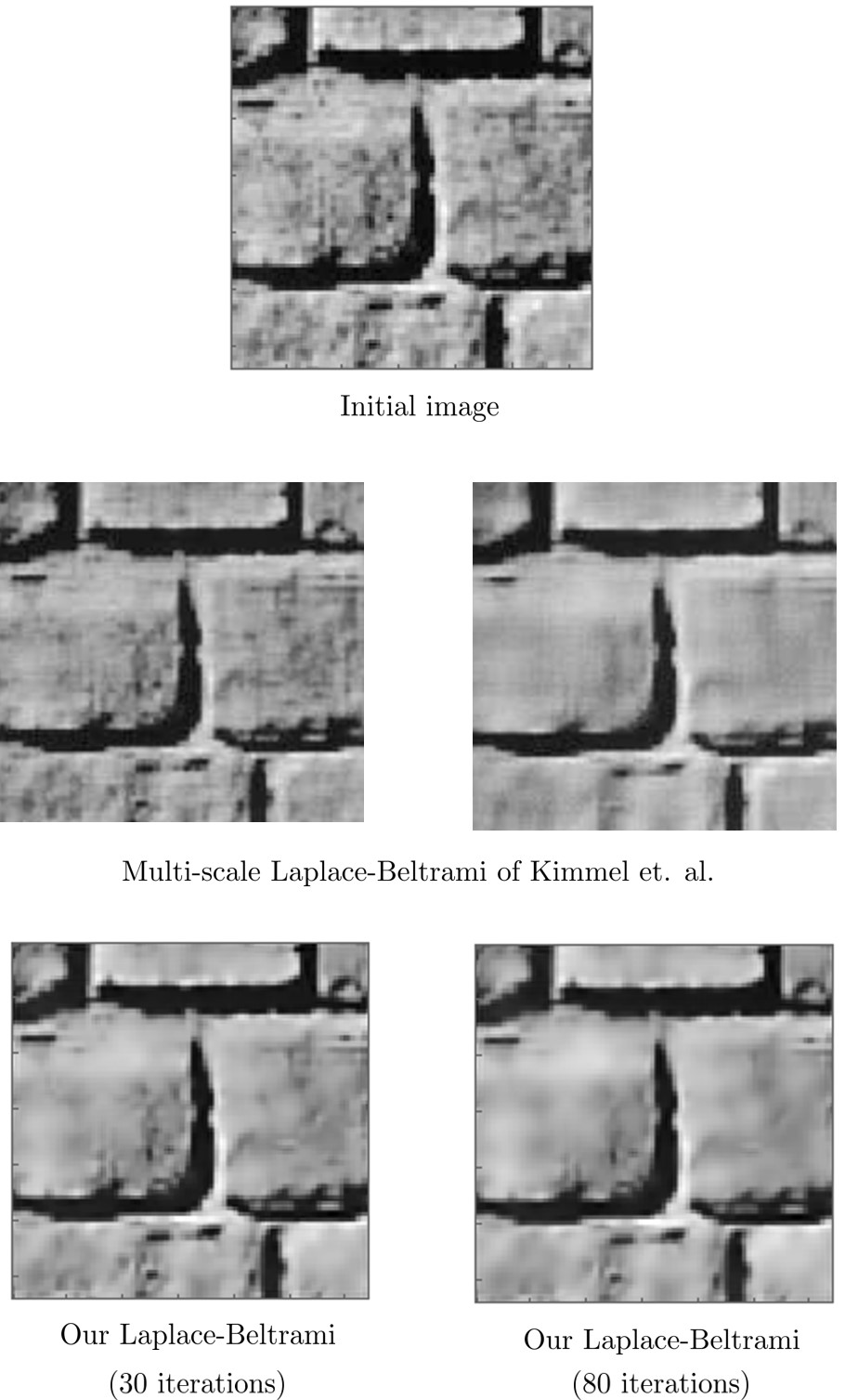}}
\caption{Top: The initial image taken from \protect\cite{kimmel2000images}. Middle: The results obtained by Kimmel et. al. \protect\cite{kimmel2000images}. Bottom: The results of our Laplace-Beltrami procedure.}
\label{fig:wallKimmelComparisonFin}
\end{figure}

We continue with Figure \ref{fig:wallKimmelComparisonFin} where we apply our procedure on a real $128\times 128$ image taken from Kimmel et. al. \cite{kimmel2000images}. In \cite{kimmel2000images} they use a multi-scale Laplace-Beltrami procedure with a fixed frequency. We use the same phase and orientation samples as in the case of Figure \ref{fig:comparisonDiffandLaplaceBeltrami2} while we employ the frequency samples $\omg\in\{ 0.5, 1,\dots, 2,2.25,\dots, 4.5, 4.625,\dots, 6 \}$ for the lifting. Here the coefficients $c_1$, $c_2$ are chosen as in the case of Figure \ref{fig:comparisonDiffandLaplaceBeltrami2}. We perform the experiments with 30 and 50 iterations.

\begin{figure}
\centerline{\includegraphics[scale=0.470,trim={0cm 0cm 0cm 0cm},clip]{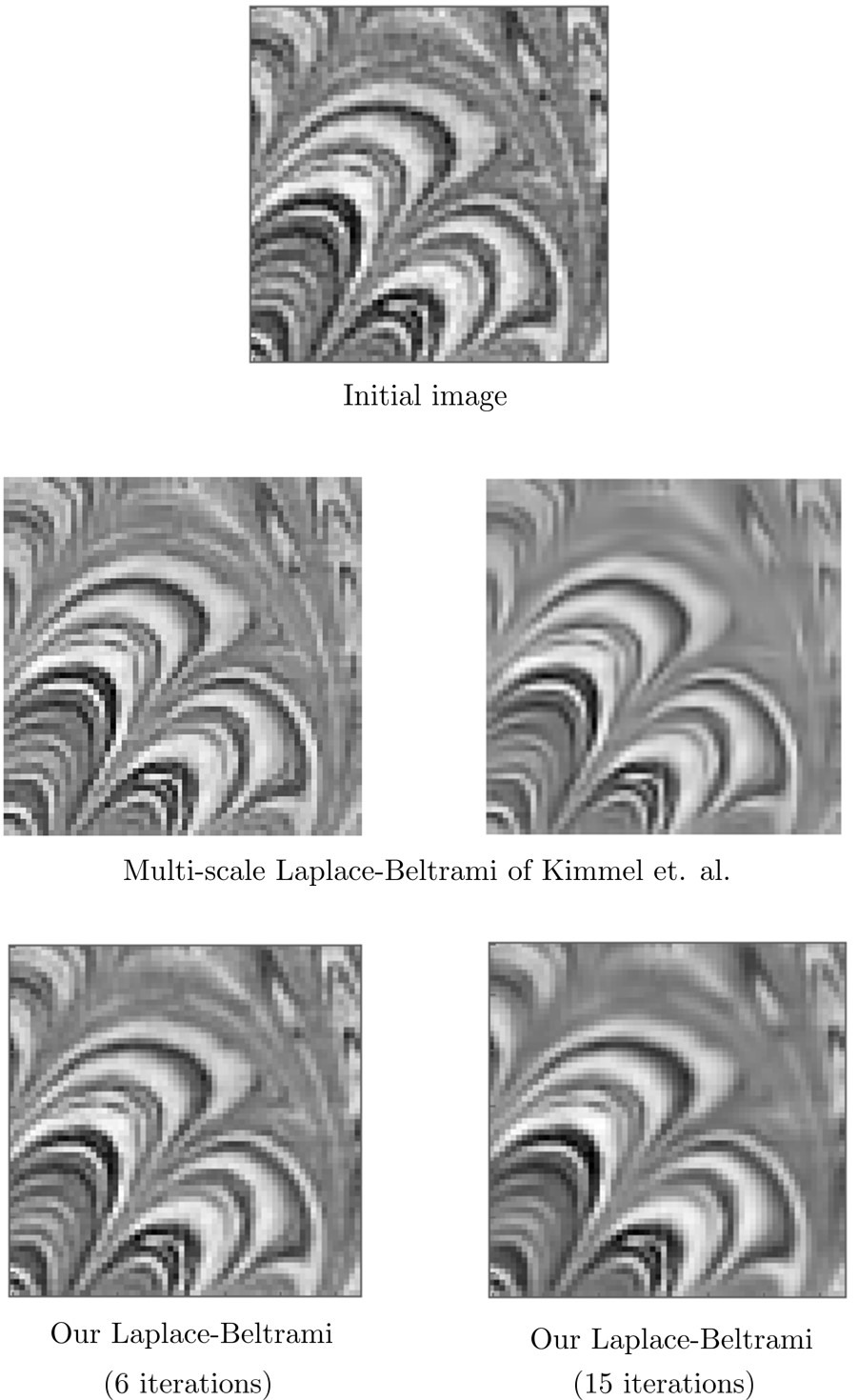}}
\caption{Top: The initial image taken from \protect\cite{kimmel2000geometry}. Middle: The results obtained by Kimmel et. al. \protect\cite{kimmel2000geometry}. Bottom: The results of our Laplace-Beltrami.}
\label{fig:kimmel_chaoticImage_comp}
\end{figure}

We show in Figure \ref{fig:kimmel_chaoticImage_comp}, the results related to our Laplace-Beltrami procedure applied on another real image, with dimensions $64\times 64$, taken from Kimmel et. al. \cite{kimmel2000images}. We use the same sampling parameters as in the previous case of Figure \ref{fig:wallKimmelComparisonFin}  for the lifting. We perform our Laplace-Beltrami procedure with 6 and 15 iterations. The results are presented together with the multi-scale Laplace-Beltrami results obtained by Kimmel et. al \cite{kimmel2000images} for a comparison. Our algorithm takes advantage of different frequencies present in images and therefore can preserve texture structures in such images as in Figure \ref{fig:kimmel_chaoticImage_comp}. Compare the elongated structures towards the right edge of the images corresponding to Kimmel et. al. \cite{kimmel2000images} (middle right) and to our procedure (bottom right).

\FloatBarrier

\section{Conclusion}

In this paper we have shown that the multi-feature selective simple cells and the associated V1 functional geometry can be modeled starting from a suitably chosen receptive profile, which was the extended Gabor function in our framework. We have derived the whole model sub-Riemannian geometry and the corresponding horizontal connectivity directly from the receptive profile. In addition to this construction of the model, we have also provided an image processing application employing our model framework: image enhancement via a sub-Riemannian Laplace-Beltrami procedure. We have provided the algorithm and its discretization explicitly as well as some experimental results. We have also mentioned that in fact, the enhancement procedure could be switched to an image inpainting procedure via a modification of the reduced metric used for the enhancement.

\section*{Funding}
G. Citti and A. Sarti are funded by the European Union’s Horizon 2020 research and innovation programme under the Marie Skłodowska-Curie grant agreement GHAIA, No 777822.

\bibliographystyle{spmpsci}
\bibliography{JMIV_Bib}

\end{document}